# Dimensionality driven enhancement of ferromagnetic superconductivity in URhGe

Daniel Braithwaite*[1], Dai Aoki[1,2], Jean-Pascal Brison[1], Jacques Flouquet[1], Georg Knebel[1], Ai Nakamura[2], Alexandre Pourret[1].

[1]Univ. Grenoble Alpes, CEA, INAC-PHELIQS, 38000 Grenoble, France
[2]Institute for Materials Research, Tohoku University, Oarai, Ibaraki 311-1313, Japan

**Abstract**

In most unconventional superconductors, like the high-$T_c$ cuprates, iron pnictides, or heavy fermion systems, superconductivity emerges in the proximity of an electronic instability. Identifying unambiguously the pairing mechanism remains nevertheless an enormous challenge. Among these systems, the orthorhombic uranium ferromagnetic superconductors have a unique position, notably because magnetic fields couple directly to ferromagnetic order, leading to the fascinating discovery of the re-emergence of superconductivity in URhGe at high field. Here we show that uniaxial stress is a remarkable tool allowing fine-tuning of the pairing strength. With a relatively small stress, the superconducting phase diagram is spectacularly modified, with a merging of the low and high field superconducting states and a significant enhancement of superconductivity. The superconducting critical temperature increases both at zero field and under field, reaching 1K, more than twice higher than at ambient pressure. The enhancement of superconductivity is directly related to a change of the magnetic dimensionality with an increase of the transverse magnetic susceptibility, demonstrating that in addition to the Ising-type longitudinal ferromagnetic fluctuations, transverse magnetic fluctuations also play an important role in the superconducting pairing.

For usual s-wave superconductors, superconductivity and ferromagnetism are antagonist states as the ferromagnetic exchange field easily destroys the superconducting pairs. Therefore the discovery of the microscopic co-existence of superconductivity and ferromagnetism in three orthorhombic uranium based heavy-fermion systems (UGe$_2$, URhGe, UCoGe)[1-6] was one of the most exciting events in recent condensed matter physics. This coexistence strongly suggests a superconducting state with triplet pairing, where the Pauli limiting mechanism is not active and the Cooper pairs can survive in the strong exchange field. The direct coupling between a static uniform field and ferromagnetism also leads to fascinating behaviour of the superconducting state under magnetic field. In these systems the pairing mechanism can actually be tuned by magnetic field, either increased[7,8] or suppressed[9], respectively when the magnetic field is applied perpendicular to (transverse field configuration) or along (longitudinal configuration) the easy magnetization axis. The most striking case is URhGe. When the magnetic field is applied along the *b*-axis of the orthorhombic crystal (perpendicular to the easy magnetisation *c*-axis), superconductivity is initially suppressed for a field of about 2 T due to the usual orbital effect. However, at higher field, superconductivity reappears in the field region from 9 - 13 T[8]. The superconducting critical temperature ($T_{SC}$) of the re-entrant phase reaches a value of 0.4 K almost twice higher than $T_{SC}$ in zero field, at a field $H_R \approx 12$ T, where a rotation of the direction of the magnetic moments occurs, from the *c* to the *b* axis[8].

Hydrostatic pressure is a powerful method to reveal unconventional superconductivity in many systems, mainly by driving them to the threshold of an electronic instability[10,11]. Previous studies[12,13] show that in URhGe, hydrostatic pressure increases the Curie temperature $T_{Curie}$, driving the system further away from its instability. Simultaneously $H_R$ was found to increase, accompanied by a collapse of zero field superconductivity at about 4 GPa and of the re-entrant superconductivity at an even lower pressure of 2 GPa[13]. Here we demonstrate that in URhGe, uniaxial stress is a much more efficient tuning tool, which succeeds to enhance the magnetic fluctuations and to drastically decrease $H_R$. We show that stress applied along the *b*-axis rapidly leads to a strong increase of the superconducting critical temperatures at zero field and at $H_R$. This increase is accompanied by a merging of the low and high field superconducting phases even before a significant effect of stress has been detected on $T_{Curie}$. The driving force for the enhancement of pairing mechanism under uniaxial stress, even in zero magnetic field, seems to be the increase of the *b*-axis susceptibility, moving the system away from the Ising-type limit. This mechanism is predicted by microscopic theories of anisotropic ferromagnetic superconductors, as arising from an enhanced coupling of the spin-polarized bands by transverse fluctuations[14].

A single crystal of URhGe was grown by the Czochralski technique. A bar shaped sample was cut with the long direction along the *b*-axis. The need to have quite a large sample meant that the sample quality was less than that of the best samples. An indication of this is the residual resistivity ratio which was RRR=10. Also at ambient pressure the sample showed a sharp but incomplete superconducting transition. The sample resistance was measured by a 4-point technique, with the current along the b-axis, in a dilution cryostat. A magnetic field of maximum 8T was applied along the *b*-axis. The sample was compressed along the *b*-axis between two sapphire anvils. The stress was applied and changed in-situ using bellows filled with helium, and determined by the force/pressure ratio of the bellows, as well as a direct measurement of force changes monitored by a piezo-electric sensor. Stress was increased up to a maximum value of 1.2 GPa. Magnetization was measured in a clamp cell adapted to a SQUID magnetometer (Quantum Design MPMS) up to 0.6 GPa, and a maximum field of 5T. The force applied at room temperature determined the stress value. More details are given in the supplemental material.

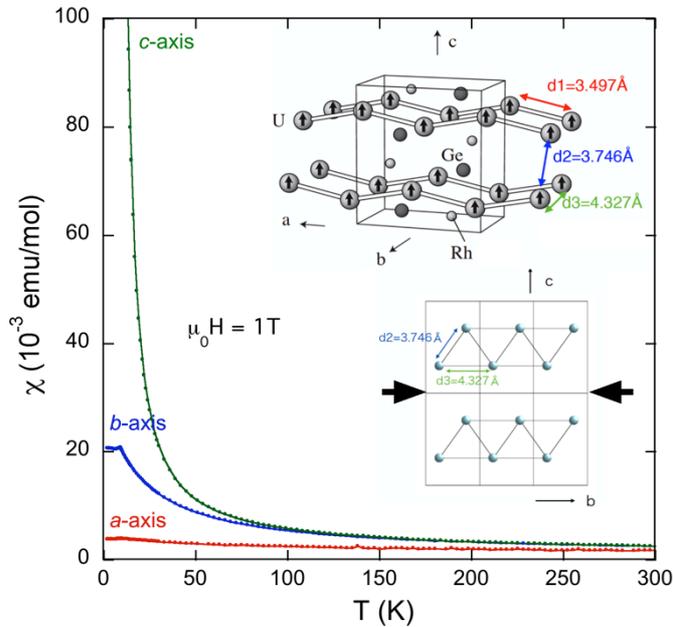

**Figure 1:** Magnetic susceptibility and orthorhombic crystal structure of URhGe showing the position of the U atoms in the *cb* plane: applied stress (black arrows) along the *b* axis reduces the distance between U atoms in *b* direction while it increases this distance in *c* direction.

In figure 1 we show the temperature dependence of the magnetic susceptibility for the 3 crystallographic directions as well as the orthorhombic crystal structure of URhGe. In the *ac* plane the U atoms form a zig-zag chain along the *a* direction with the magnetic moments aligned in the *c* direction. The U atoms are slightly displaced in the *bc* plane, and in the *ab* plane, they form distorted hexagons, slightly elongated along the *b* axis. Hence the structural anisotropy in the *bc* plane is rather small. While the *a*-axis is the hardest magnetic axis at all temperatures, the magnetic susceptibilities along the *b* and *c* axis are almost the same above 50 K. Thus, the system seems to select the easy magnetization *c*-axis and form a ferromagnetic ground state, only when a coherent state with a Fermi surface of renormalized heavy quasiparticles develops. The low energy scale governing the emergence of the *cb* anisotropy is certainly a key element driving the response of URhGe to a magnetic field $H//b$: a strong increases of the magnetization along *b*, together with a sharp suppression of the *c* axis component of the magnetization at $H_R$, leading to re-entrant superconductivity accompanied by an abrupt Fermi surface reconstruction[15,16].

Generally, when stress is applied along one crystallographic axis, the unit cell is compressed in this direction, but expands in the other directions. From thermal expansion measurements[17] according to the Ehrenfest relation, applying uniaxial stress along the *b*-axis of URhGe would tend to reduce $T_{Curie}$ with an expected slope of -1.6 K/GPa. We also expect that, as the characteristic magnetic energy scales are lowered, the rotation of the moments from the *c* to the *b* axis will occur at lower field. The first goal is therefore to determine the relations between $T_{Curie}$, $H_R$, and $T_{SC}$, the latter both in zero field, and at $H_R$.

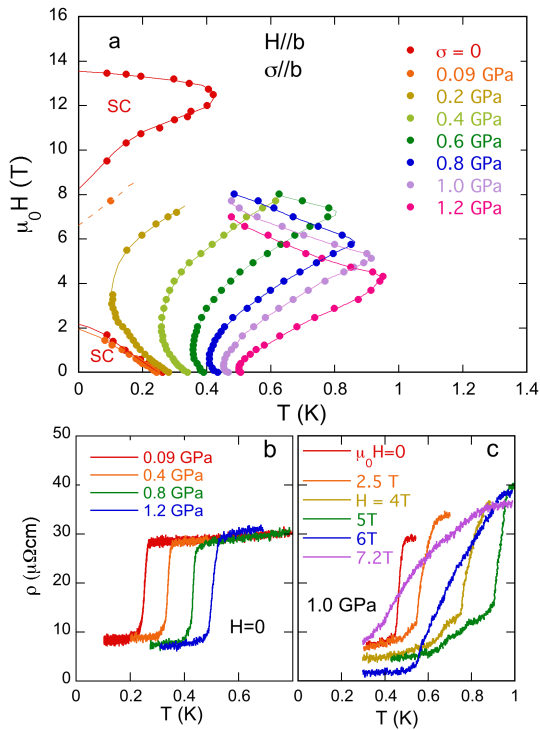

**Figure 2:** $(T,H)$ phase diagram of superconductivity for the transverse ($H//b$) configuration. 2(a) shows the phase diagram for different values of stress, together with a previous zero pressure measurement[7]. It is obtained with a criterion of 50% of the total resistance drop. Lines are guides to the eye. Figures 2(b) and 2(c) show the superconducting transitions at different values of stress and field respectively. See supplemental material[18] for a discussion on the shape of the transitions.

Figure 2 shows the $(T, H)$ superconducting phase diagram in the transverse field configuration ($H//b$) for different values of uniaxial stress applied along the *b* axis. We have defined $H_R$, as the field of the maximum $T_{SC}$ of the high field phase; it corresponds also to the maximum of the normal state magnetoresistance (see supplemental material). We find two main effects. First of all, $H_R$ decreases strongly: the high field superconducting pocket moves to lower fields and merges with the low field superconducting phase already at 0.2 GPa. Secondly, at the same time, superconductivity is significantly enhanced, both at zero field, and even more strongly at $H_R$.

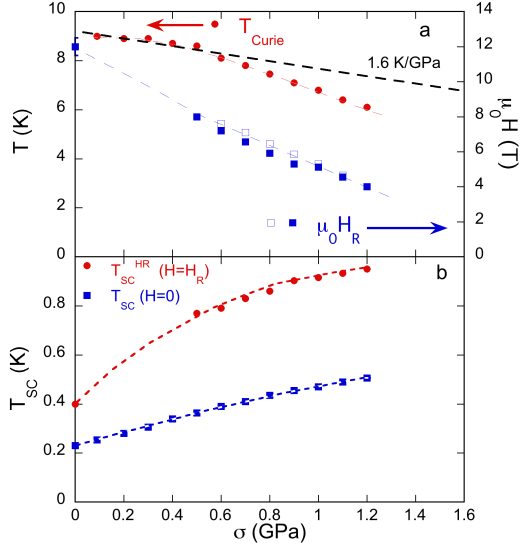

**Figure 3:** (a) Stress dependence of the parameters $T_{Curie}$ and field $H_R$ where the rotation of the moments occurs. $H_R$ was determined by the maximum of $T_{SC}$ (full symbols) and by the peak in the normal state magnetoresistance (open symbols; see supplemental material[18]). As $H_R$ for $\sigma=0$ was not measured on this sample, a typical value of 12T was taken, the error bar shows the possible uncertainty, mainly due to the precision of the sample alignment with the field. The dashed black line shows the expected initial slope for $T_{Curie}$ of 1.6 K/GPa from the thermal expansion measurements. (b) stress dependence of the superconducting critical temperature $T_{SC}^{(H=0)}$ at zero field (red circles) and at the maximum value $T_{SC}^{HR}$ obtained for $H=H_R$. Lines are guides for the eyes.

In figure 3 we show the stress dependence of the different parameters, $T_{Curie}$, $H_R$, $T_{SC}^{(H=0)}$ and the maximum $T_{SC}^{HR}$. As expected from the thermal expansion results, $T_{Curie}$ decreases with stress. This decrease is very weak below 0.5 GPa, but quantitatively, the stress dependence of $T_{Curie}$ is compatible with a slope of ~ -1.6K/GPa, deduced from thermal expansion. This decrease is more pronounced above 0.6 GPa. In contrast, $H_R$ is extremely sensitive to uniaxial stress. $H_R$ could not be determined below for stress below 0.6 GPa as $H_R$ is larger than 8T, the maximum field we could apply for these measurements, however the stress dependence of $H_R$ is close to linear over the whole measured range. At 0.6 GPa $H_R$ has decreased to less than 8T and it is reduced to 4T at 1.2 GPa, the maximum stress achieved. This rapid decrease of $H_R$ is accompanied by a significant enhancement of the superconducting critical temperatures: $T_{SC}^{(H=0)}$ and $T_{SC}^{HR}$ both increase continuously, $T_{SC}^{(H=0)}$ reaching 0.5K and $T_{SC}^{HR}$ almost attaining 1K at 1.2GPa. But like for $H_R$, their strong initial variations under stress contrast with the weak decrease of $T_{Curie}$. And this variation is not altered above 0.6GPa, when $T_{Curie}$ is decreasing faster. So qualitatively, the evolution of the critical field $H_R$ and of the superconducting transitions seems *not* driven by the mere evolution of $T_{Curie}$ under stress.

A hint for the factor which may be most influenced by stress comes from the strong decrease of $H_R$: the relation between the high-field superconducting state, and the rotation of the moments at $H_R$ was established early on[8]. So qualitatively, the lowering of $H_R$ means that under stress, a smaller field is required to align the moments along the *b*-axis. In other words: the anisotropy between the *c*-axis (the easy magnetization axis) and the *b*-axis is getting smaller. Quantitatively, it was noticed that the rotation occurs when the magnetization in the *b*-direction approaches the spontaneous zero-field magnetization $M_c = 0.4$ $\mu_B$ in the easy *c*-direction[7,17]. This means that if $H_R$ occurs at lower fields, the susceptibility along the *b*-axis, $\chi_b = \partial M/\partial H$, should increase, with a proportionality of $1/\chi_b \approx H_R$. Figure 4(a) shows our magnetization measurements under uniaxial stress: a significant and rapid increase of the susceptibility $\chi_b$ with increasing stress is found. In Fig. 4(b) we show the relative change of the different parameters versus stress. The variations of $1/\chi_b$ and $H_R$ are quite similar, as expected, whereas the relative change of $T_{Curie}$ is much weaker. It is also clear that $T_{SC}$ is correlated to $1/\chi_b$ and $H_R$ rather than to $T_{Curie}$. This extreme sensitivity of $H_R$ to stress, and the fact that $T_{Curie}$ is much less so, points to strongly anisotropic magnetocrystalline effects. The situation is quite different with hydrostatic pressure, where up to 0.84 GPa, $T_{Curie}$ and $H_R$ were found to increase with similar relative changes (about 10% and 16% respectively[13]). This can be qualitatively understood by considering the crystal structure of URhGe (see (ure 1). Applying stress along the *b*-axis will reduce the nearest neighbour distance of U atoms in the *b*-direction, while the nearest neighbour distance along the *c*-axis will increase, leading through the variation of the exchange integrals to an increase of the susceptibility along the *b*-axes and a decrease of $T_{Curie}$. It will also reduce the distortion of the hexagons leading to a more isotropic *bc* plane.

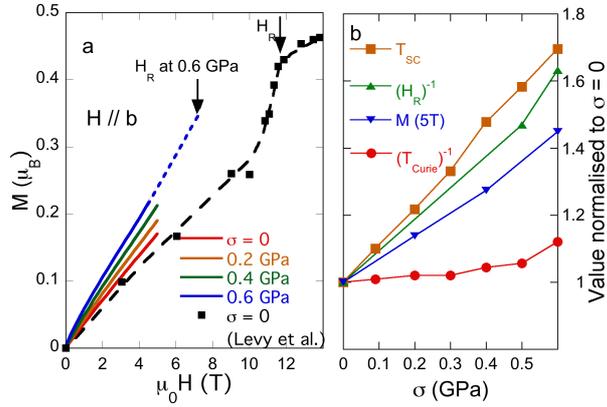

**Figure 4:** Magnetization (a) of URhGe under uniaxial stress for $H//b$ and $\sigma//b$, compared to the ambient pressure data of Levy et al.[8]. The arrows show the value of $H_R$ at ambient pressure and $\sigma = 0.6$ GPa. Figure 4(b) shows that the normalized changes of $\chi$ (taken as $M$ at 5T), $1/H_R$, and and $T_{SC}$ are very similar, and seem uncorrelated with those of $1/T_{Curie}$. Lines are guides for the eyes.

The other spectacular result is that with increasing stress, superconductivity is strongly enhanced, both at zero field and at $H_R$, with the maximum $T_{SC}^{HR}$ more than doubling between ambient pressure and 1.2 GPa. Of course as $H_R$ decreases, the maximum $T_{SC}^{HR}$ occurs at a lower field, so the orbital pair-breaking effect of the field is weaker and will naturally lead to a higher $T_{SC}$.

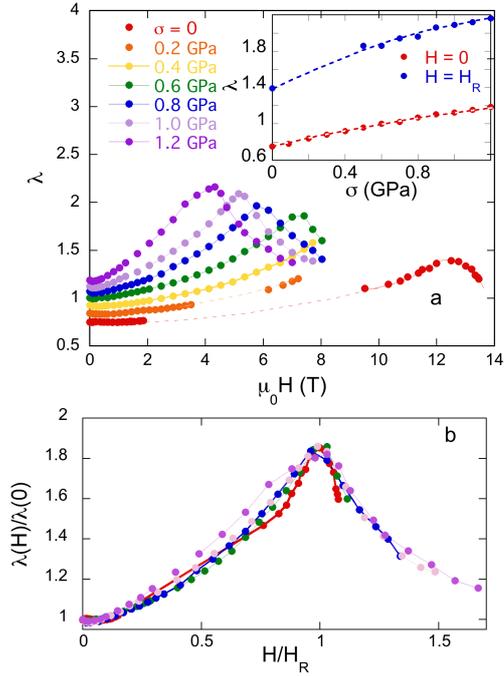

**Figure 5:** analysis of $H_{C2}$. 5(a): Field dependence of the strong coupling parameter $\lambda$ determined from the fits of the upper critical field $H_{C2}$ for different applied stresses. See supplemental material [18] for a description of the procedure. The red points are taken from ref[8] where the zero pressure analysis was performed. The insert shows the pressure dependence of $\lambda$ at $H=0$ and $H=H_R$. Figure 5(b) shows the normalized values; the enhancement of $\lambda$ at $H_R$ is constant at all values of stress. Lines are guides for the eye.

To eliminate the influence of the reduced orbital effect when $H_R$ occurs at lower field, and better quantify the reinforcement of superconductivity, we have analysed the superconducting upper critical field ($H_{C2}$) curves using a strong-coupling model[19], following the approach developed in Wu et al.[9]. The hypothesis is that all changes of $T_{SC}$ and of the effective mass $m^*$ are controlled by a field and stress-dependent strong-coupling parameter $\lambda$ (see supplemental material for more details[18]). The results are shown in figure 5 together with those obtained from the analysis of the ambient pressure data. As stress is increased, the values of $\lambda$ at zero field and at $H_R$ clearly increase (figure 5(a) inset), showing that the pairing strength is significantly enhanced as $H_R$ moves to lower values. However, as shown in figure 5b, the enhancement of $\lambda$ between zero field and $H_R$, plotted versus $H/H_R$, appears to be independent of stress. In other words the value of λ (and therefore $T_{SC}$) at $H_R$ under stress scales with its zero field value.

This increase of $T_{SC}$ with uniaxial stress at zero field, while $T_{Curie}$ is almost constant, is quite remarkable. We have shown that it matches the observed decrease of $H_R$, and the proportional increase of the susceptibility along the *b*-axis, which can be seen as a weakening of the uniaxial anisotropy of URhGe. This is completely at odds with initial theoretical studies predicting that Ising type ferromagnets would be more favourable to equal spin pairing (ESP) p-wave superconductors due to the suppression of the "pair-breaking" transverse magnetic fluctuations[20-22].

More recent work has shown that this is not the case in anisotropic (orthorhombic) systems[14]. Taking into account the coupling between the two components ($\Delta_{\uparrow\uparrow}$ and $\Delta_{\downarrow\downarrow}$) of the ESP p-wave order parameter, a weak-coupling expression for $T_{SC}$ very similar to that for a multiband superconductor is derived[14]. Intra-band couplings $g_{1\uparrow,\downarrow}$, are proportional to the susceptibility $\chi_c$ at the Fermi wave-vector along the (easy) c-axis, and to the respective averaged density of states for spin ↑, ↓ bands. Inter-band couplings $g_{2\uparrow,\downarrow}$ depend mainly on the same density of states and on $(\chi_b - \chi_a)$. In the isotropic systems considered in the early theories, $g_{2\uparrow,\downarrow} = 0$. For an orthorhombic anisotropy, $g_{2\uparrow,\downarrow} \neq 0$, and $T_{SC}$ is expressed as[14]:

$$T_{SC} \propto \exp\left(-\frac{1}{g}\right)$$

$$g = \frac{g_{1\uparrow} + g_{1\downarrow}}{2} + \sqrt{\frac{(g_{1\uparrow} - g_{1\downarrow})^2}{4} + g_{2\uparrow}g_{2\downarrow}}$$

$g_{2\uparrow}g_{2\downarrow} \propto (\chi_b - \chi_a)^2$ is a positive term, showing directly that the anisotropy of the transverse susceptibilities increases $T_{SC}$. We claim that this is the reason why stress is so effective in boosting $T_{SC}$: enhancing $\chi_b$ drives the system from a "1D" towards a "2D" magnetic anisotropy, increasing the coupling between the superconducting order parameters of the opposite spin bands, which enhances $T_{SC}$ as in any two-band superconductor. This mechanism seems much more efficient than the simultaneous weak decrease of $T_{Curie}$. The situation for the re-entrant phase at $H_R$ is more complex. Close to this field, NMR studies have shown that both the longitudinal and transverse ferromagnetic fluctuations are strongly enhanced [23,24]. Our results strongly suggest that both types of fluctuations might contribute to reinforce superconductivity at $H_R$.

Therefore, the zero field results yield the most solid experimental argument demonstrating that in ferromagnetic systems, a large transverse susceptibility allowing the presence of both longitudinal and transverse ferromagnetic fluctuations is more effective for superconductivity, than a strong Ising character. If this is well understood as arising from "multiband superconductivity", a detailed microscopic theory of the magnetic properties of URhGe is still missing and the details of the electronic structure is a difficult and open question: we hope that this work will motivate theoretical efforts towards a more quantitative understanding of the interplay between ferromagnetism and superconductivity in uranium systems. Experimentally, these results should stimulate new investigations on URhGe and might also guide explorations to find other ferromagnetic superconductors. The immediate challenge is to succeed in performing experiments at higher stress: extrapolating our results we find that $H_R$ should be tuned to zero at a critical value of stress of about 1.7 GPa. At this point, URhGe might switch from an easy *c*-axis to a *b*-axis ferromagnetic state. What will happen then to superconductivity at this point is a completely open question.

**Acknowledgements**

We thank V. Mineev and W. Knafo for stimulating discussion. Support for this work was provided by the ANR (PRINCESS), CEFIPRA (ExtremeSpinLadder) and ERC (NewHeavyFermion) projects. One of us (D.A.) acknowledges support from KAKENKI, ICC-IMR, and REIMEI programs

Supplemental material

**Determination of $T_{SC}$ and $T_{Curie}$**

The single crystal of URhGe was grown by the Czochralski technique in a tetra-arc furnace. The crystal was oriented, and cut to the dimensions 2.45 x 1.34 x 1.20 mm$^3$, where the long direction is the *b*-axis. Resistivity was measured in a dilution refrigerator with current 100 μA applied along the b-axis at 18.7 Hz and lock-in detection. The sample was placed in a uniaxial stress cell and compressed between 2 sapphire anvils. The stress was applied and changed in-situ using bellows filled with helium, and determined by the force/pressure ratio of the bellows, as well as a direct measurement of force changes monitored by a piezo-electric sensor. Stress was increased up to a maximum value of 1.2 GPa, where one contact was lost. Measurements were made at constant magnetic field. A calibrated RuO$_2$ thermometer on the cell measured the sample temperature. The heater is placed close to the cell and the cell is sufficiently decoupled from the mixing chamber to allow slow (2K/hour) temperature sweeps up to 4K without excessive heating of the mixing chamber or the bellows. To determine T$_{Curie}$ the cell was heated very rapidly (in less than 1 minute) to about 12K, to avoid excessive heating of the mixing chamber and the liquid helium in the bellows. The heater was then switched off and measurement was performed as the cell was allowed to cool naturally. This cooling was also rapid (typically 1K/min) so some error on the temperature is expected. This is estimated to be less than 0.5K and should be a systematic error so the variation of $T_{Curie}$ with stress is quite reliable. High temperature curves are shown in figure S1a. $T_{Curie}$ appears as a clear kink at low values of stress, becoming a more broad rounded feature at higher stress. We have determined the values of $T_{Curie}$ by taking the second derivative of the resistivity temperature dependence. We find an initial decrease of $T_{Curie}$ with a slope of about -1K/GPa. From the thermal expansion data we evaluate the expected initial slope through the Ehrenfest relation and find -1.6 K/GPa.

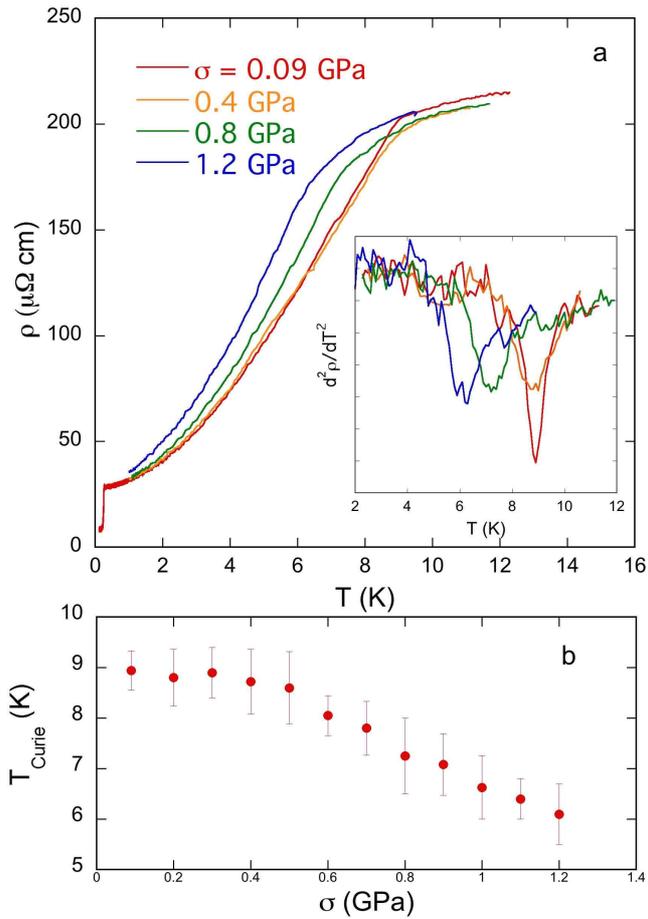

**Figure S1:** S1(a): High temperature curves ρ(T) for the determination of $T_{Curie}$. The inset shows the second derivative which was used to place the transition. The stress dependence of is shown in figure S1(b) together with a typical transition width taken as the width of the negative peak in $d^2\rho/dT^2$ at half maximum height.

**The superconducting transitions and phase diagram**

The resistive superconducting transition was not complete (i.e. zero resistance was never achieved) even at zero stress and field. This is probably because of the insufficient quality of the sample. The superconducting transition temperature, $T_{SC}$, was determined by taking the criterion of 50% resistance change between the onset and low temperature. This produces a relatively symmetric superconducting pocket centred on $H_R$. However the transitions above and below $H_R$ are quite different as can be seen in figure S2a. Below $H_R$ the transition stays rather sharp, though a "foot" starts to grow on approaching $H_R$. Above $H_R$ the transition becomes much broader. The symmetry created by the 50% criterion is therefore somewhat artificial. As zero resistance is never achieved we cannot use this as a criterion, however to illustrate the effect in figure S2b we show the phase diagram obtained using the criterion of 90% of the total resistance drop. Then superconductivity in fact disappears rather rapidly above $H_R$ which probably reflects better the intrinsic behaviour of bulk superconductivity. This is consistent with the behaviour at ambient pressure and is probably related to the 1$^{st}$ order nature of the transition at $H_R$. The change of criterion does not significantly affect the value of $H_R$ so the results presented in the article of the article are robust and independent of the chosen criterion.

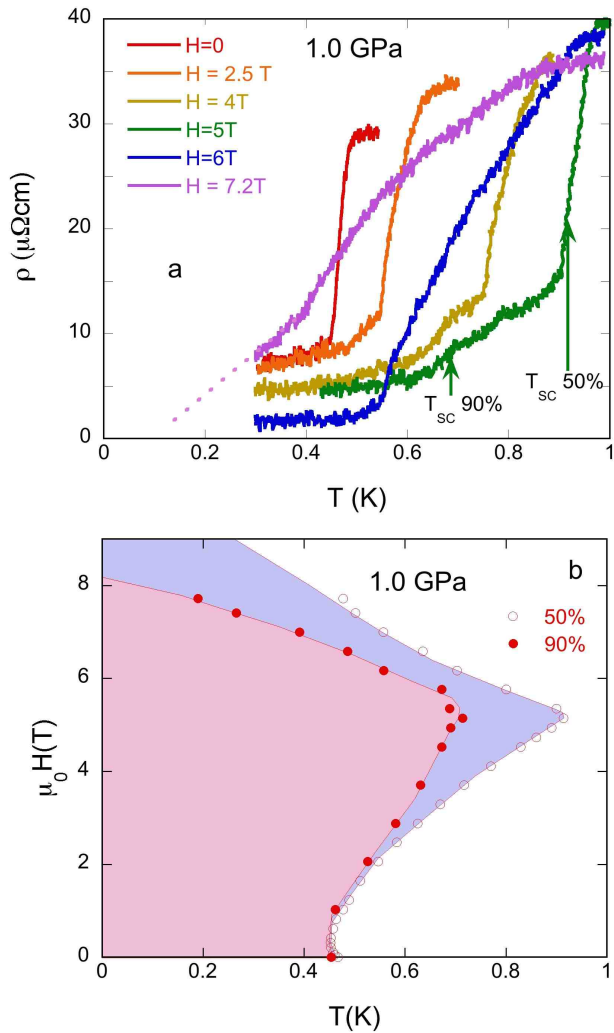

**Figure S2:** Influence of the superconducting transition criterion on the phase diagram. S2(a) : Evolution of the superconducting transition with field with stress of 1 GPa. Figure S2(b) shows the influence of the criterion on the resulting phase diagram implying that bulk superconductivity probably disappears rapidly for fields greater than $H_R$.

**Normal state resistivity**

The temperature dependence of the resistivity was measured accurately up to 4K at zero field, and at some selected field values. At all values of stress we performed 2 field sweeps, at 1K and 3K. Magnetoresistance curves are shown in figure S3a showing that $H_R$ also appears as a peak in the normal state resistivity. This is important, as superconductivity measured from the resistivity can be a non-intrinsic effect due for example to filamentary or surface superconductivity, whereas the normal state resistivity is a bulk measurement. The fact that we find a peak in the normal state resistivity shows that the phenomenon at $H_R$ is indeed a bulk effect, so superconductivity is almost certainly bulk too. A Fermi liquid behaviour ($\rho=\rho_0+AT^2$) was found at all values of field and stress. The stress dependence of A at zero field is shown in figure S3(b)

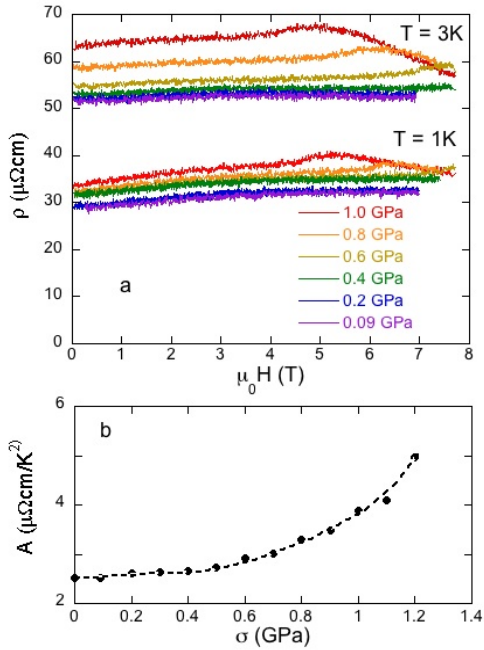

**Figure S3:** Normal state resistivity. S3(a): Magnetoresistance in the normal state at 1K and 3K showing the existence of an anomaly at $H_R$. Figure S3(b) shows the stress dependence of the prefactor A of the Fermi-liquid temperature dependence of the normal state resistivity.

The rather small initial stress dependence of A implies that the effective mass is also only changing weakly (usually A is proportional to $m^{*2}$). The absolute values of A may be however not very reliable, due to stress inhomogeneity. Nevertheless the stress dependence of A is compatible with the variation of $T_{Curie}$. This means that the mechanism responsible for enhancing superconductivity, which we believe is related to the increase of the b-axis susceptibility, has little influence on A. This is not necessarily surprising as the sensitivity of the transport coefficients to the microscopic excitations is complex, and it depends strongly on the nature of these excitations. We can also evaluate the initial relative change of the effective mass from the slope of the thermal expansion data. This also gives an expected weak initial increase of $m^*$ (about 10% / GPa), compatible with the normal state resistivity data, and with the initial insensitivity to stress.

**Magnetization measurements under stress**

A hydrostatic pressure cell designed for use in a commercial magnetometer (Quantum Design MPMS) was adapted for uniaxial stress measurements. The sample was pressed between 2 ceramic pistons. CuBe Belleville washers were inserted to maintain an almost constant force despite the thermal contractions of the different parts of the cell. We checked that no significant change of force was found between 300K and 77K, and we assume this is true down to the lowest temperature. We estimated that if a change occurs it will be an increase in stress on cooling with a maximum value of 0.05 GPa. The sample was first measured without the cell, then re-measured at zero stress in the cell. This allowed to subtract the cell contribution. Measurements were carried out up to 0.6 GPa. Because magnetization measures the whole volume of the sample, the question of the stress homogeneity is even more crucial than for resistivity. In uniaxial stress measurements it is known that especially close to the ends of the sample stress is not homogeneous. The magnetization measurements therefore probably underestimate the effect of stress, and the real increase of $\chi$ is almost certainly larger, and may compare even better with the decrease of $H_R$.

**Determination of the stress and field dependence of the strong-coupling parameter $\lambda$.**

A realistic description of the pairing mechanism in superconductors involves the momentum and frequency dependence of the interaction between quasiparticles and excitations responsible for the pairing; it involves as well as the complete spectrum of these excitations. However, simplified models capturing most of the new physics induced by strong coupling effects depend mainly on a strong coupling parameter $\lambda$, which is a weighted integral over frequency of the density of interactions and excitation spectrum. This parameter is the equivalent of the product of the density of states by interaction potential in BCS theory, and it controls notably the superconducting transition temperature $T_{SC}$. It controls also normal state quantities like the contribution of the pairing interaction to the renormalization of the quasi particles effective mass, or equivalently, of their Fermi velocities: $m^*/m=1+\lambda$. Our main assumption is that most of the evolution of the superconducting properties under stress, and the anomalous behavior under magnetic field (re-entrant phase et $H_R$) is governed only by a change of $\lambda$. The physical reason is that it is a very sensitive parameter: in the weak-coupling limit, $T_{SC}$ depends exponentially on $1/\lambda$, whereas it depends only linearly on the average excitation frequency (the Debye frequency for electron-phonon interactions). Other parameters like the screened Coulomb potential $\mu^*$ are governed by large energy scales and logarithmic cut-off, so they are expected to change very little in the small stress of field range applied. Using a minimal strong coupling model [19] to calculate the $T_{SC}$ dependence on $\lambda$ and the corresponding $H_{c2}$, (taking the dependence of the Fermi velocity into account), we could extract, as explained in reference [9] how $\lambda$ should vary under stress and field, once the value for $\lambda$ (H=0, $\sigma$ =0) is properly chosen.